\documentclass[pre,preprint]{revtex4}
\begin{document}
\draft
\bibliographystyle{revtex4}
\titlepage
\title{The coexistence of $p$-wave  spin triplet superconductivity
and itinerant ferromagnetism}
\author{Hong-Biao Zhang}
\email{hbzhang@eyou.com} \affiliation{Institute of Theoretical
Physics, Northeast Normal University, Changchun, Jilin 130024,
P.R.China} \affiliation {Theoretical Physics Division, Nankai
Institute of Mathematics, Nankai University, and Liuhui Center for
Applied Mathematics, Tianjin 300071, P.R.China}
\author{Mo-Lin Ge}
\email{geml@nankai.edu.cn} \affiliation {Theoretical Physics
Division, Nankai Institute of Mathematics, Nankai University, and
Liuhui Center for Applied Mathematics, Tianjin 300071, P.R.China}
\author{Li-Jun Tian}
\affiliation {Theoretical Physics Division, Nankai Institute of
Mathematics, Nankai University, and Liuhui Center for Applied
Mathematics, Tianjin 300071, P.R.China}
\begin{abstract}
A model for coexistence of p-wave spin-triplet superconductivity
(SC) and itinerant ferromagnetism (FM) is presented. The
Hamiltonian can be diagonalized by using the so(5) algebraic
coherent state. We obtain the coupling equations of the magnetic
exchange energy and superconducting gaps through the double-time
Green function. It is found that the  ferromagnetisation gives
rise to the phase transitions of p-wave superconducting states or
superfluid of $^{3}He$.
\end{abstract}
\maketitle

\baselineskip=30pt
\section{Introduction}
\indent Since Ginzburg\cite{fmgin} pointed out a possibility of
the coexistence between ferromagnetism (FM) and superconductivity
(SC) for the magnetization less than the thermodynamic critical
field, many experimental investigations were made, for example,
for impurity ferromagnetism in a superconductor\cite{fmmat}. As
well-known, the coexistence between antiferromagnetism (AFM) and
SC looks easy to be realized and observed in several
compounds\cite{fmmaple} because AFM moments spatially averaged
over the SC coherent length vanish, but difficult for FM case. The
rare exceptions are rare-earth ternary compounds $HoMo_6S_8$ and
$ErRh_4B_4$ where in narrow region just below the Curie
temperature $T_{FM}$ the coexistence of FM and SC was
attained\cite{fmhk}. When the rare-earth $4f$ moments completely
align at lower temperature, SC is wiped out by a strong internal
field. It seems that so far not known SC can fully sustain such a
large molecular field. As for the theoretical developments,
starting by Anderson and Suhl\cite{fmanderson} there had been
discussing the possible coexistence\cite{fmfulde,fmlark,fmblount}.
The possibility of a finite momentum paring state coexisting with
the long range FM order was presented and SC in metals with a
spin-exchange field produced by FM aligned impurities was
considered. In such a SC ferromagnet there are two kinds of
electrons respectively, responsible for FM and SC. One is
localized electrons forming a FM background in metal indirect
exchange coupling through itinerant electrons, the other forms
Cooper pairs due to the effective attractive interaction.
Recently, the theoretical works show the coexistence of weak
itinerant FM with s-wave SC\cite{fmblag,fmbedell}.

On the other hand, the recent discovery of the coexistence FM and
SC in $VGe_2$\cite{fmsaxena,fmbauer}, and subsequently in
$ZrZn_2$\cite{fmple} and $URhGe$\cite{fmaoki}, have shown clearly
that the long awaited spin-triplet SC state is realized in the
nature. In these experiments it looks to have a weak and itinerant
nature of electrons involved in both FM and SC. This naturally
renews our interest in the relationship between FM and SC for
p-wave SC (or superfluid) because a spin-triplet may form the
Anderson-Morel state with combination of
$|\uparrow\uparrow\rangle$ and
$|\downarrow\downarrow\rangle$\cite{fmleggett1,fmlee}.

In this paper we will consider a general model for the coexistence
between p- SC and itinerant FM. As the known result there is
$so(5)$ structure in p-SC\cite{fmzhang2,murakami} that is formed
by two $su(2)'s$ not commuting with each other where one $su(2)$
describes the attractive BCS interaction and the other for the
usual spin operators as well as other 4 generators relating to the
transitions. Motivated by Ref.\cite{fmbedell}, we write the
Hamiltonian in two part $H=H_{SC}+H_{FM}$ where $H_{SC}$ is the BW
type of p-SC Hamiltonian i.e. $H_{SC}=\sum\limits_{{\bf
k},\alpha}\epsilon_{\bf k} a^{\dag}_{{\bf k}\alpha}a_{{\bf
k}\alpha} +\frac{1}{2}\sum\limits_{{\bf k},{\bf k}',\alpha,\beta}$
$V_{{\bf k}{\bf k}'}a^{\dag}_{-{\bf k}'\alpha} a^{\dag}_{{\bf
k}'\beta}a_{{\bf k}\beta}a_{-{\bf k}\alpha}$. After taking the
mean-field approximation we have the reduced Hamiltonian of the
coexistence of p-SC and itinerant FM:
\begin{eqnarray}
\label{fm2.1} &&H=\sum_{\bf k}\epsilon_{\bf k} (a^{\dag}_{{\bf
k}\uparrow}a_{{\bf k}\uparrow}+a^{\dag}_{{\bf k}\downarrow}a_{{\bf
k}\downarrow}) -\sum_{\bf k}(\Delta_{\alpha\beta}({\bf
k})a^{\dag}_{{\bf k}\beta}a^{\dag}_{-{\bf k}\alpha}+H.c.)
\nonumber \\
&&\;\;\;\;\;\;\;\;+\sum_{\bf k}\Delta_{\alpha\beta}({\bf
k})<a^{\dag}_{{\bf k}\beta}a^{\dag}_{-{\bf k}\alpha}>
-\frac{JM}{2}\sum_{\bf k}(a^{\dag}_{{\bf k}\downarrow}a_{{\bf
k}\downarrow} -a^{\dag}_{{\bf k}\uparrow}a_{{\bf
k}\uparrow})+\frac{1}{2}JM^{2}
\end{eqnarray}
\begin{equation}
\label{mm} M=\frac{1}{2}\sum_{\bf k}(<a^{\dag}_{{\bf
k}\downarrow}a_{{\bf k}\downarrow}>
-<a^{\dag}_{{\bf k}\uparrow}a_{{\bf k}\uparrow}>) \\
\end{equation}
\begin{equation}
\label{fmd} \Delta_{\alpha\beta}({\bf k})=-\frac{1}{2}\sum_{{\bf
k}'} V_{{\bf k}{\bf k}'}<a_{{\bf k}'\alpha}a_{-{\bf k}'\beta}>
\end{equation}
where $\epsilon_{\bf k}=\frac{p^2}{2m^*}-\mu$ is the band energy
measured from the chemical potential, and for $p$-wave attraction
pair interaction potential $V_{{\bf k}{\bf k}'}=-3V_{1}(k,k'){\bf
n}\cdot{{\bf n}'}$ (${\bf n}=\frac{\bf k}{k}$). $<\cdots>$
represents the thermodynamic average, $M$ defines the
magnetization of the system, and $\Delta_{\alpha\beta}$ is the
superconducting gap. We note that the eq.(\ref{fm2.1}) is made up
of $p$-wave SC terms and FM term. The two constant terms in
eq.(\ref{fm2.1}) result from the mean-field approximation, the
first constant term comes from the BCS interaction and the second
one from the exchange coupling. Here the magnetization defined in
eq.({\ref{mm}}) arises from a spontaneously breaking of spin
rotation symmetry of the itinerant electrons, which is different
from a paramagnetic response to a magnetic field caused by
localized spins. Therefore, both the gap and the magnetic exchange
energy are determined by eq.(\ref{fmd}) and eq.(\ref{mm})
self-consistently, unlike in the conventional metal with magnetic
impurities where the exchange energy is considered as an external
parameter.

Next we diagonalize the reduced Hamiltonian (\ref{fm2.1}) using
Lie algebra $so(5)$ coherent state approach. The generators of Lie
algebra $so(5)$ is expressed as\cite{fmzhang2,fmzhang3}
\begin{equation}
\label{matrix} I_{ab}({\bf k})=\left(
\begin{array}{ccccc}
0&&&&\\
-\frac{1}{2}(T^{\dag}_{1}({\bf k}) + T_{1} ({\bf k})) &0&&&\\
-\frac{1}{2}(T^{\dag}_{2}({\bf k}) + T_{2}({\bf k})) &
-F_{3}({\bf k}) &0&&\\
-\frac{1}{2}(T^{\dag}_{3}({\bf k}) + T_{3} ({\bf k})) &
F_{2}({\bf k}) &-F_{1}({\bf k}) &0&\\
{Q}({\bf k})&\frac{1}{2i}(T_{1}({\bf k}) - T^{\dag}_{1}({\bf k}))&
\frac{1}{2i}(T_{2}({\bf k}) - T^{\dag}_{2}({\bf k}))&
\frac{1}{2i}(T_{3}({\bf k})-T^{\dag}_{3}({\bf k}))&0\\
\end{array}
\right)
\end{equation}
where ${\bf F}({\bf k})=\frac{1}{2}[{\bf S}({\bf k}) +{\bf
S}(-{\bf k})]$, ${Q}({\bf k})=\frac{1}{2}[S_{0}({\bf k})
+S_{0}(-{\bf k})-2]$ and ${S}_{i}({\bf k})={a}^{\dag}_{{\bf
k}\alpha}{({\sigma_{i}})_{\alpha\beta}}{a_{{\bf k}\beta}}$ and
$T_{i}({\bf k})=a_{-{\bf
k}\alpha}(\sigma_{2}\sigma_{i})_{\alpha\beta}{a_{{\bf k}\beta}}$
as well as their conjugates ($i=0,1,2,3$, $\sigma_{0}={\bf 1}$ and
summation over the repeated $\alpha$ and $\beta$) is given
in\cite{fmzhang2,fmzhang3}. It can be proved that $I_{ab}$ obey
the following commutation relation:
 $$
 [{I}_{ab}({\bf k}), {I}_{cd}({\bf k}')]=-i\delta{({\bf k} -{\bf k}')}
 (\delta_{ac}{I}_{bd}({\bf k}) + \delta_{bd} {I}_{ac}({\bf k})
  -\delta_{ad}{I}_{bc}({\bf k}) - \delta_{bc}{I}_{ad}({\bf k}) )
 $$
and $I_{ab}({\bf k})$ = $-I_{ba}({\bf k})$ ($a,b=1,2,3,4,5$) is
antisymmetric matrix element.

 Therefore, we can rewrite eq.(\ref{fm2.1}) in terms
of the generators of the Lie algebra $so(5)$ as follows:
 \begin{equation}
 \label{fm2.2}
 H=\sum_{{\bf k}}H({\bf k})-E_{0}
 \end{equation}
 \begin{equation}
 \label{fm2.3}
 H({\bf k})=\epsilon_{{\bf k}}Q({\bf k})
 +{\bf\Delta}({\bf k})\cdot{\bf T}^{\dag}({\bf k})
 +{\bf\Delta}^{\dag}({\bf k})\cdot{\bf T}({\bf k})+JMS_{3}({\bf k})
 \end{equation}
 \begin{equation}
 E_{0}=\sum\limits_{{\bf k}}[\epsilon_{{\bf k}}-{\bf\Delta}({\bf k})\cdot
 <{\bf T}^{\dag}({\bf k})>]+\frac{1}{2}JM^{2}
 \end{equation}
 \begin{equation}
 {\bf\Delta}({\bf k})=\frac{1}{4}\sum\limits_{{\bf k}'}V_{{\bf k}{\bf k}'}
 <{\bf T}({{\bf k}'})>
 \end{equation}
 \begin{equation}
M=\frac{1}{2}\sum\limits_{{\bf k}}<F_{3}({\bf k})>
 \end{equation}
Here we emphasized that the set
$\Lambda=\{\frac{i}{\sqrt{2}}T_{3}({{\bf k}}),
\frac{-i}{\sqrt{2}}{T^{\dag}_{3}}({{\bf k}}),Q({{\bf k}})\}$ i.e.
$\{-i\sqrt{2}\pi_{z},i\sqrt{2}\pi^{\dag}_{z},-Q \}$ in
\cite{fmzhang2} forms the quasi-spin $\Lambda$. $\Lambda$ does not
commute with spin operators ${\bf S}({\bf k})$ that give rise to
$T^{\dag}_{\pm}({\bf k})$ and $T_{\pm}({\bf k})$ which are beyond
two $su(2)$ and the total set forms $so(5)$. In order to perform
the diagonalization of eq.(\ref{fm2.3}) we introduce the unitary
transformation $U(\xi_{{\bf k}})$) such that $U^{\dag}(\xi_{{\bf
k}})H({\bf k})U(\xi_{{\bf k}}) =E_{{\bf k}\downarrow}n_{{\bf
k}\downarrow} +E_{{\bf k}\uparrow}n_{{\bf k}\uparrow}$ becomes
diagonal for each given momentum ${\bf k}$. Following the general
strategy \cite{fmzhang3} we introduce the $so(5)$-coherent
operators:
\begin{equation}
U(\xi_{{\bf k}})=\exp\{\xi_{{\bf k}}[{\bf d}({\bf n})\cdot{\bf
T}^{\dag}({\bf k})]-H.c.\}
\end{equation}
where $\xi_{\bf k}$ is called the coherent parameter, ${\bf
d}({\bf n})=(\sin{\psi_{\bf k}}\cos{\phi_{\bf k}}, \sin{\psi_{\bf
k}}\sin{\phi_{\bf k}},\cos{\psi_{\bf k}})$ corresponding to the
direction of zero spin projection, $\psi_{\bf k}$ and $\phi_{\bf
k}$ are angles in spin space for a given momentum ${\bf k}$.
Taking the commutation relations for $so(5)$ into account after
lengthy but elementary calculations we derive the following two
different solutions for ${\bf d}({\bf n})$.

1. When $\cos{\psi_{{\bf k}}}=0$,  the direction of the pair
orbital angular momentum ${\bf L}$ is perpendicular to the
direction of zero spin projection ${\bf d}$, the energy is split
into:
\begin{equation}
E_{{\bf k}\uparrow}=(1+\frac{JM}{\epsilon_{\bf k}})E_{\bf k}
\end{equation}
\begin{equation}
E_{{\bf k}\downarrow}=(1-\frac{JM}{\epsilon_{\bf k}})E_{\bf k}
\end{equation}
\begin{equation}
E_{\bf k}=\sqrt{\epsilon^2_{\bf
k}+4[\frac{|\Delta_{\uparrow\uparrow}({\bf k})|^2}
{(1+\frac{JM}{2\epsilon_{\bf k}})^2}
+\frac{|\Delta_{\downarrow\downarrow}({\bf
k})|^2}{(1-\frac{JM}{2\epsilon_{\bf k}})^2}]}
\end{equation}
This result exhibits that the SC energy is split by the
magnetization $M$. For finite temperature $T$, making use of the
double-time Green function  we find $M$ and the non-vanishing
components of $\Delta_{\alpha\beta}({\bf k})$ :
\begin{equation}
\label{m} M=\frac{1}{4}\sum\limits_{\bf k}\frac{\epsilon_{\bf
k}}{E_{\bf k}} [\tanh\frac{\beta}{2}(1+\frac{JM}{2\epsilon_{\bf
k}})E_{{\bf k}\uparrow}
-\tanh\frac{\beta}{2}(1-\frac{JM}{2\epsilon_{\bf k}})E_{{\bf
k}\downarrow}]
\end{equation}
\begin{equation}
\label{del1} \Delta_{\uparrow\uparrow}({\bf k})
=-\frac{1}{2}\sum\limits_{{\bf k}'}V_{{\bf k}{\bf k}'}
\frac{\Delta_{\uparrow\uparrow}({\bf
k}')}{(\frac{JM}{2\epsilon_{{\bf k}'}}+1)E_{{\bf k}'}}
\tanh\frac{{\beta}E_{{\bf k}'\uparrow}}{2}
\end{equation}
\begin{equation}
\label{del2} \Delta_{\downarrow\downarrow}({\bf k})
=-\frac{1}{2}\sum\limits_{{\bf k}'}V_{{\bf k}{\bf
k}'}\frac{\Delta_{\downarrow\downarrow}({\bf
k}')}{(\frac{JM}{2\epsilon_{{\bf k}'}}-1)E_{{\bf k}'}}
\tanh\frac{{\beta}E_{{\bf k}'\downarrow}}{2}
\end{equation}
From the  gap equations (\ref{del1}) and (\ref{del2}) we read that
under finite temperature, FM and $p$-wave SC may coexist in the
$p$-wave equal spin pairing state(ABM state), i.e. $\Psi_{AM}\sim
e^{\phi}{\sin{2|\xi|}}(|\uparrow\uparrow>+e^{\chi}|\downarrow\downarrow>)$.
But if choosing $p^{\pm}_{F}=\sqrt{m^{*}(2\mu\pm JM)}$, then the
phase $A$ of the equal spin pairing state will turn into phase
$A_{1}$ (with only spin up pairing $|\uparrow\uparrow>$ ) and
phase $A_{2}$  (with only spin down pairing
$|\downarrow\downarrow>$) respectively.  At temperature $T=0$, let
us distinguish two cases for eqs.(\ref{m})-(\ref{del2}):

(a). If $JM=2\epsilon_{\bf k}$ or
$p^{+}_{F}=\sqrt{m^{*}(2\mu+JM)}$, then the above consistent
equations reduce to
\begin{equation}
M=\frac{1}{4}\sum_{\bf k}\frac{\epsilon_{\bf k}}
{\sqrt{\epsilon^2_{\bf k}+|\Delta_{\uparrow\uparrow}({\bf k})|^2}}
\end{equation}
\begin{equation}
\Delta_{\uparrow\uparrow}({\bf k})=-\frac{1}{4}\sum\limits_{{\bf
k}'}V_{{\bf k}{\bf k}'}\frac{\Delta_{\uparrow\uparrow}({\bf k}')}
{\sqrt{\epsilon^2_{{\bf k}'} +|\Delta_{\uparrow\uparrow}({\bf
k}')|^2}}
\end{equation}
\begin{equation}
\Delta_{\downarrow\downarrow}({\bf k})=0
\end{equation}
that indicates the gap equation described by the Anderson-Morel
state with  only $|\uparrow\uparrow>$ spin pairs. As a physical
consequence the existence of FM may turn the p-wave EPS (phase $A$
in ${}^3He$) into phase $A_{1}$   with the only state
$|\uparrow\uparrow>$.

(b). if $JM=-2\epsilon_{\bf k}$ or
$p^{-}_{F}=\sqrt{m^{*}(2\mu-JM)}$, then we have
$\Delta_{\uparrow\uparrow}({\bf k})=0$ and all the
$\uparrow\uparrow$ in (16) and (17) are replaced by
$\downarrow\downarrow$. The gap equation is described by ABM state
with  only $|\downarrow\downarrow>$ spin-down pairs, i.e. the
phase $A$ can be turned into phase $A_{2}$   with only
$|\downarrow\downarrow>$. Therefore,  the coexistence of FM and
$p$-wave SC gives rise to the phase transitions from phase $A$
 to phase $A_{1}$ or $A_{2}$ . Such a phase
transition may be  observed in the coexistence of FM and SC for
p-SC and ${}^3He$ superfluid.

2. When $\sin{\psi_{\bf k}}=0$, the direction of the pair orbital
angular momentum ${\bf L}$ is parallel to the direction of zero
spin projection ${\bf d}$ and  the energy is  split into
\begin{equation}
E_{{\bf k}\uparrow}=E_{\bf k}+\frac{JM}{2}
\end{equation}
\begin{equation}
E_{{\bf k}\downarrow}=E_{\bf k}-\frac{JM}{2}
\end{equation}
\begin{equation}
E_{\bf k}=\sqrt{\epsilon^{2}_{\bf
k}+4|\Delta_{\uparrow\downarrow}({\bf k})|^2}
\end{equation}
that is the same as the s-wave case\cite{fmbedell} which exhibits
a two-fold Zeeman splitting effect in the itinerant FM. For finite
temperature $T$ by making use of the double-time Green function
the $M$ and the non-vanishing components of
$\Delta_{\alpha\beta}({\bf k})$
$(\Delta_{\uparrow\downarrow}=\Delta_{\downarrow\uparrow})$ can be
calculated:
\begin{equation}
M=\frac{1}{4}\sum\limits_{\bf k}(\tanh\frac{{\beta}E_{{\bf
k}\uparrow}}{2} -\tanh\frac{{\beta}E_{{\bf k}\downarrow}}{2})
\end{equation}
\begin{equation}
\Delta_{\uparrow\downarrow}({\bf k})=-\frac{1}{4}\sum\limits_{{\bf
k}'}V_{{\bf k}{\bf k}'} \frac{\Delta_{\uparrow\downarrow}({\bf
k}')}{E_{{\bf k}'}}(\tanh\frac{{\beta}E_{{\bf k}'\uparrow}}{2}
+\tanh\frac{{\beta}E_{{\bf k}'\downarrow}}{2})
\end{equation}
The above gap equation  can be described by the opposite spin
pairing state of $p$-wave SC, i.e. $\Psi_{AM}\sim
e^{\phi}\sin{|\xi|}[\frac{1}{\sqrt{2}}(|\uparrow\downarrow>
+|\downarrow\uparrow>)]$.

At $T=0$, we obtain the same relations as given by
\cite{fmbedell}:
\begin{equation}
p^{\pm}_{F}=\sqrt{2m^{*}\mu\pm
m^{*}\sqrt{(JM)^{2}-16|\Delta_{\uparrow\downarrow}|^{2}}}
\end{equation}
\begin{equation}
\label{m1} M=\frac{1}{4}\int
\frac{d^{3}p}{(2\pi)^{3}}=\frac{1}{12{\pi}^{2}}
[(p^{+}_{F})^{3}-(p^{+}_{F})^{3}]
\end{equation}
\begin{equation}
\label{del3} \Delta_{\uparrow\downarrow}({\bf
k})=-\frac{1}{4}\sum\limits_{{\bf k}'}V_{{\bf k}{\bf k}'}
\frac{\Delta_{\uparrow\downarrow}({{\bf k}'})}
{\sqrt{\epsilon^2_{{\bf k}'}+4|\Delta_{\uparrow\downarrow}({\bf
k}'})|^2}
\end{equation}
The only solutions of eqs.(\ref{m1}) and (\ref{del3}) for the
coexistence of SC and FM occur for
$JM>4|\Delta_{\uparrow\downarrow}|$ that can be  discussed as the
same as $s$-wave case given  in Ref.\cite{fmbedell}. Therefore,
the opposite spin pairing triplet state behaviors very similar to
the s-wave singlet state in the present of exchange splitting.
This leads to the conclusion that the effect of the different
exchange splitting of SC state is determined by whether the state
contains OSP .(The spin singlet and the opposite spin pairing
state of the  $p$-wave triplet are, by definition, OSP states.) or
ESP, and the FM can make transition of  the  phase $A$ of SC state
to the phase $A_{1}$  or phase $A_{2}$  under ESP state.

This work is in part supported by the NSF of China.


\noindent

\begin{thebibliography}{99}
\bibitem{fmgin} V. L. Ginzberg, Sov. Phys. JETP {\bf 4} (1957) 153.
\bibitem{fmmat} B. T. Matthias et al., Phys. Rev. Lett. {\bf 1}
(1959) 92.
\bibitem{fmmaple} M. B. Maple and F. Fisher, Superconductivity in
Terary Compounds I and II, edited by Springer-Verlag, Berlin,
1982.
\bibitem{fmhk} K. Machida and H. Nakanishi, Phys. Rev. B {\bf 30}
(1984) 122.
\bibitem{fmanderson} P. W. Anderson and H. Suhl, Phys. Rev. {\bf 116}
(1959) 898.
\bibitem{fmfulde} P. Fulde and R. A. Ferrell, Phys.
Rev. {\bf 135} (1964) A550.
\bibitem{fmlark} A. I. Larkin and Yu. N. Ovchinnikov, Zh. Eksp.
Teoz. Fiz. {\bf 47} (1964) 1136. (Sov. Phys. JETP {\bf 20} (1975)
762.)
\bibitem{fmblount} E. I. Blount and C. Varma, Phys. Rev.
Lett. {\bf 42} (1979) 1079.
\bibitem{fmblag} K. B. Blagoev et al., Phys. Rev. Lett. {\bf 82}
(1999) 133.
\bibitem{fmbedell} N. I. Karchev, K. B. Blagoev, K. S.
Bedell and P. B. Littlewood, Phys. Rev. Lett. {\bf 86} (2001) 846.
\bibitem{fmsaxena} S. Saxena et al.,  Nature {\bf 406} (2000) 587.
\bibitem{fmbauer} E. Bauer, R. Dickey, V. Zapf and M. Maple, J.
Phys. Cond. Matt. {\bf 13} (2001) L759.
\bibitem{fmple} C. Pfleiderer, M. Uhlarz, et al., Nature (London) {\bf
412} (2001) 58.
\bibitem{fmaoki} D. Aoki, A. Huxley et al., Nature (London)
{\bf 413} (2001) 613.
\bibitem{fmleggett1} A. J. Leggett, Ann.Phys., NY {\bf{85}} (1974) 11-55.
\bibitem{fmlee} D. M. Lee, Rev. Mod. Phys. {\bf 69} (1997)
645-664.
\bibitem{fmzhang2} D. J. Scalapinp, S. C. Zhang and W. Hanke, Phys. Rev. B {\bf 58} (1998)
443; S. C. Zhang, Science {\bf 275} (1997) 1089.
\bibitem{murakami} S. Murakami and N. Nagaosa, M. Sigrist, Phys.
Rev. Lett. {\bf 82} (1999) 2939-2942.
\bibitem{fmzhang3} H. B. Zhang, M. L. Ge and K. Xue, J.
Phys. A: Math. Gen. {\bf 35} (2002) L7-L11.

\end{thebibliography}
\end{document}